\documentclass[twocolumn]{aastex631}
\pdfoutput=1

\usepackage{graphicx}
\usepackage{booktabs} 
\usepackage{threeparttable}
\begin{document}

\title{Driving factors behind multiple populations}

\author{Ruoyun Huang}
\address{School of Physics and Astronomy, Sun Yat-sen University, Zhuhai 519082, China}
\address{CSST Science Center for the Guangdong-Hong Kong-Macau Greater Bay Area, Zhuhai 519082, China}

\author{Baitian Tang}
 \email{tangbt@mail.sysu.edu.cn}
\address{School of Physics and Astronomy, Sun Yat-sen University, Zhuhai 519082, China}
\address{CSST Science Center for the Guangdong-Hong Kong-Macau Greater Bay Area, Zhuhai 519082, China}

\author{Chengyuan Li}
\address{School of Physics and Astronomy, Sun Yat-sen University, Zhuhai 519082, China}
\address{CSST Science Center for the Guangdong-Hong Kong-Macau Greater Bay Area, Zhuhai 519082, China}

\author{Doug Geisler}
\address{Departamento de Astronomia, Universidad de Concepcion, Concepcion, Chile}
\address{Instituto de Investigación Multidisciplinario en Ciencia y Tecnología, Universidad de La Serena, La Serena, Chile}
\address{Departamento de Astronomía, Facultad de Ciencias, Universidad de La Serena, La Serena, Chile}

\author{Mario Mateo}
\address{Department of Astronomy, University of Michigan, Ann Arbor, MI 48109, USA }

\author{Ying-Yi Song}
\address{David A. Dunlap Department of Astronomy \& Astrophysics, University of Toronto, Toronto, ON M5S 3H4, Canada}
\address{Dunlap Institute for Astronomy \& Astrophysics, University of Toronto, Toronto, ON M5S 3H4, Canada}

\author{Holger Baumgardt}
\address{School of Mathematics and Physics, The University of Queensland, Australia}

\author{ Julio A. Carballo-Bello}
\address{Instituto de Alta Investigaci\'on, Universidad de Tarapac\'a, Arica, Chile}

\author{Yue Wang }
\address{CAS Key Laboratory of Optical Astronomy, National Astronomical Observatories, Chinese Academy of Sciences, Beijing 100101, China}

\author{Jundan Nie }
\address{CAS Key Laboratory of Optical Astronomy, National Astronomical Observatories, Chinese Academy of Sciences, Beijing 100101, China}

\author{Bruno Dias }
\address{Instituto de Astrofísica, Facultad de Ciencias Exactas, Universidad Andres Bello, 7591538, Las Condes, Santiago, Chile}

\author{Jos\'e G. Fern\'andez-Trincado}
\address{Instituto de Astronom\'ia, Universidad Cat\'olica del Norte, Antofagasta, Chile}

\begin{abstract}

Star clusters were historically considered simple stellar populations, with all stars sharing the same age and initial chemical composition. However, the presence of chemical anomalies in globular clusters (GCs), called multiple stellar populations (MPs), has challenged star formation theories in dense environments. Literature studies show that mass, metallicity, and age are likely controlling parameters for the manifestation of MPs. Identifying the limit between clusters with/without MPs in physical parameter space is crucial to reveal the driving mechanism behind their presence. In this study, we look for MP signals in Whiting 1, traditionally considered a young GC.
Using the Magellan telescope, we obtained low-resolution spectra within $\rm \lambda\lambda = 3850-5500 \r{A}$ for eight giants of Whiting 1. We measured the C and N abundances from the CN and CH spectral indices. C and N abundances have variations comparable with their measurement errors ($\sim0.1$ dex), suggesting that MPs are absent from Whiting 1. Combining these findings with literature studies, we propose a limit in the metallicity vs. cluster compactness index parameter space, which relatively clearly separates star clusters with/without MPs (GCs/open clusters). This limit is physically motivated. On a larger scale, the galactic environment determines cluster compactness and metallicity, leading to metal-rich, diffuse, old clusters formed ex situ. Our proposed limit also impacts our understanding of the formation of the Sagittarius dwarf galaxy: star clusters formed after the first starburst (age$\lesssim 8-10$ Gyr). These clusters are simple stellar populations because the enriched galactic environment is no longer suitable for MP formation.

\end{abstract}


\section{Introduction} \label{sec:intro}
Historically, all stars within a cluster are expected to form in a single burst from the same molecular cloud, which is assumed to be well mixed. Therefore, they are treated as a simple stellar population (SSP) with the same age and initial chemistry. This assumption is generally confirmed in open clusters (OCs) \citep{ref1, ref2}. However, globular clusters (GCs) show distinctive features: Photometric and spectroscopic data indicate the presence of star-to-star chemical variations, which are known as multiple stellar populations (MPs,\citealt{ref3,ref4,ref5}). First-generation (FG) stars have similar metallicity, while the so-called second-generation (SG) stars typically show enhanced He, N, Na, (and Al) abundances but depleted C, O, (and Mg) abundances. Spectroscopic observations reveal the presence of Na--O and C--N anticorrelations in most GCs, whereas an Mg--Al anticorrelation appears preferentially in metal-poor or massive GCs \citep{ref3,ref6,ref7}. The MP phenomenon is not exclusive to GCs in the Milky Way; it is also found in GCs of other local group galaxies, such as M31 \citep{ref8}, the Magellanic Clouds \citep{ref9,ref10}, Sagittarius (Sgr,\citealt{ref11}), and Fornax \citep{ref12}.
Because MPs are exclusively found in GCs but not in OCs, \cite{ref13} defined the $bona fide$ GCs as ``stellar aggregates showing the Na--O anticorrelation,'' or, in other words, MPs. Finding key parameters that efficiently separate GCs from OCs, or more precisely clusters with and without MPs, would considerably impact not only the scenarios of star cluster formation but also the details of galaxy formation.

With an increasing number of GCs investigated, metallicity, mass, and cluster age have been proposed to be the driving factors behind MPs \citep{ref13,ref14,ref15,ref16,ref17}. Among these attributes, cluster mass is undoubtedly the most important: (1) The ratio of FG stars to SG stars and the complexity of their chemical compositions are correlated with cluster mass, with generally higher mass clusters having a larger proportion of SG stars \citep{ref18}; (2) according to the self-enrichment hypothesis, the most widely accepted scenario for MPs \citep{ref19,ref20}, the enriched gas from FG stars is expelled and mixed with the interstellar medium. Therefore, SG stars that subsequently form from this mixture should exhibit distinct chemical patterns or anomalies indicative of enrichment. To maintain the enriched gas that is essential for SG star formation, the cluster must have a sufficiently large gravitational potential well. In other words, less massive clusters may not be able to retain the chemical signature associated with the MP phenomenon.\,\,\,
To determine this possible mass threshold, several low-mass, old clusters were investigated. For example, \cite{ref15} used high-resolution spectroscopy to discover MPs in NGC 6535 (with a mass of approximately $\rm 2.2 \times 10^4 M_\odot$);  \cite{ref17} found substantial C and N inhomogeneities among the red giant members of Palomar 13 (with a mass of approximately $\rm 3 \times 10^3 M_\odot$), which is the lowest mass GC with MPs found to date

To this end, Whiting 1 with a mass of $\rm 2\times10^3 M_\odot$ \citep{ref21} is an object worthy of dedicated inspection. Discovered by \cite{ref22}, it was originally considered an OC based on its diffuse morphology.
However, given its intermediate age ($\rm 6.5^{+1.0}_{-0.5}Gyr$), relatively low metallicity ($\rm [Fe/H] = -0.65$), and high Galactic latitude (($l, b$) = (161.6$^{\circ}$, --60.6$^{\circ}$), ($\alpha,\delta$) = (02$^h$02$^m$57$^s$, --03$^o$15$^{'}$10$^{''}$)), Whiting 1 was subsequently dubbed a GC \citep{ref23, ref24}.

On the basis of its location and measured kinematic properties, Whiting 1 coincides with the tidal arms of Sgr in the models of  \cite{ref25}, which was later confirmed by Gaia measurements \citep{ref26}.
Embedded in the Sgr tidal arms, Whiting 1 has developed interesting extratidal features.  \cite{ref27} found that the structure of Whiting 1 was elongated, consistent with the orbit of the Sagittarius dwarf spheroidal galaxy (Sgr dSph).
Possible leading and trailing tails on both sides of the cluster were reported by  \cite{ref28}. These tails align with the orbital direction of Sgr dSph and the cluster itself, suggesting that this debris is most likely the remnants of Whiting 1 that were stripped away by the Milky Way.

In this study, we search for the presence of the MP phenomenon in Whiting 1 using low-resolution UV-blue spectra obtained by the Magellan Telescope.
In Section 2, we describe how we performed the observations and extracted the one-dimensional spectra. We then measure the stellar parameters and C and N abundances in Section 3. We discuss our results in a broader context in Section 4 and summarize them in Section 5.

\section{OBSERVATION AND DATA REDUCTION} \label{sec:style}

\begin{figure}[b]
   \centering
   \includegraphics[width=0.46\textwidth]{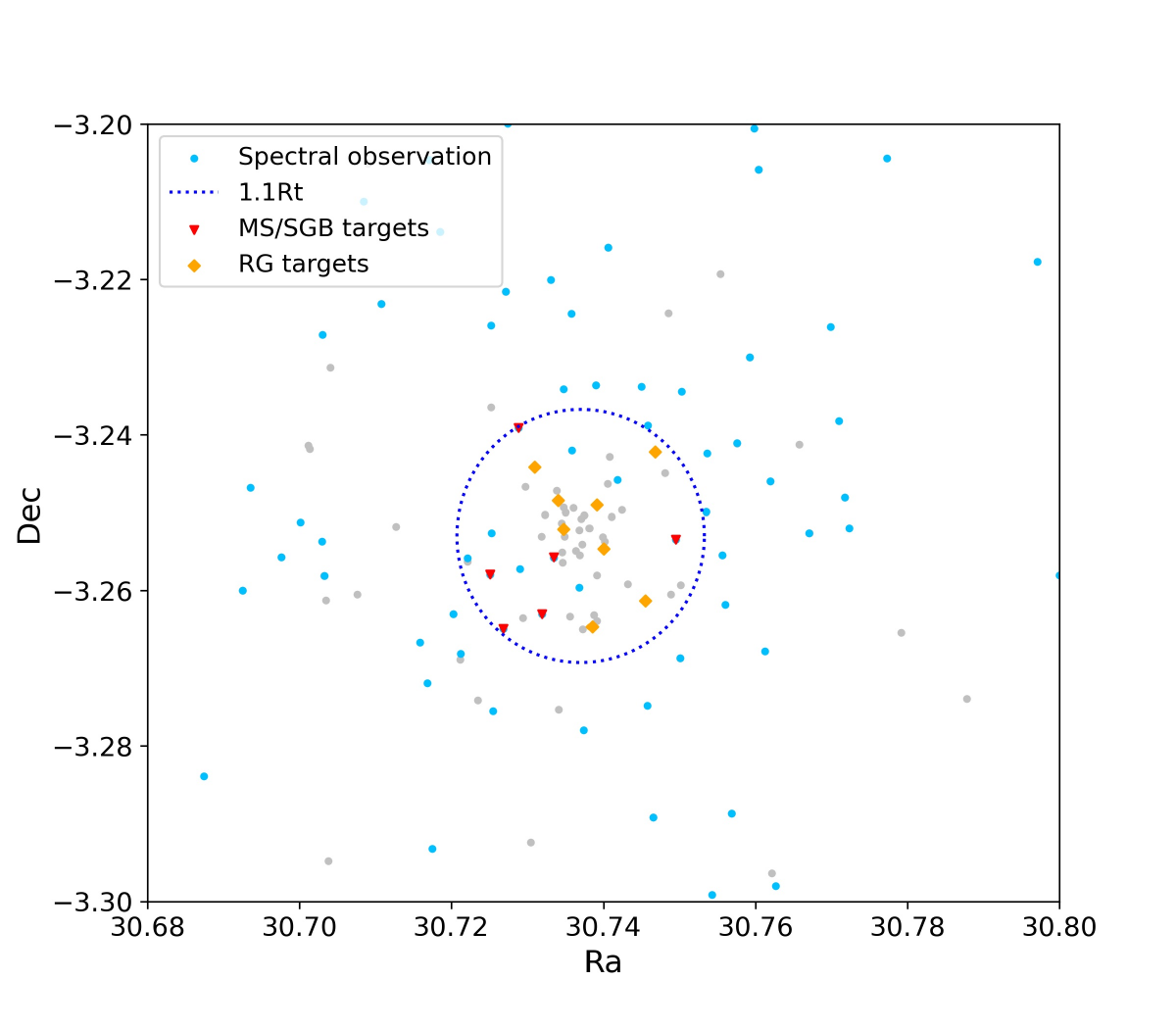}
      \caption{ Stars around Whiting 1. Markers of different colors (gray and others) represent stars with Gmag brighter than 21 mag (from Gaia DR3). The orange and red dots denote the final selected Whiting 1 targets, while light blue dots indicate additional stars observed in our spectroscopic measurements. The blue line represents 1.1 times the tidal radius of Whiting 1. The orange and red dots denote the final selected targets.}

         \label{position}
   \end{figure}

The observations were performed using M2FS (Michigan Magellan Fiber System, \citealt{ref29}) at the Magellan/Clay Telescope (Las Campanas Observatory, Chile). Because our primary focus was the molecular features (CN3839 and CH4300) around 4000 \AA, we picked the ``BK7'' filter with a 600 line/mm grating and 2$\times$2 binning for the blue and red arms. This setup gives low-resolution ($R\sim3000$) spectra with $\lambda\lambda = 3850-5500$ \AA. Each plate covers a circular area with a diameter of $29.2^\prime$, in which 128 fibers are available for each arm. We used Gaia DR2 photometry and astrometry (in 2018) to exclude foreground stars with large parallaxes. To avoid fiber collisions, two targets must be separated by at least 13'' center-to-center. We selected Whiting 1 targets, giving higher priority to brighter stars. Lower priority was assigned to fainter stars that could also be observed when free fibers were available. Figure \ref{position} shows the observed targets around Whiting 1.
Fortunately, Whiting 1 is surrounded by the Sgr tidal stream, providing targets of high scientific value. Finally, we filled 256 fibers with 248 scientific objects, while the remaining eight fibers were used to observe the sky background. The observations were executed on November 30 and December 1, 2018, with a total exposure time of 7 h\footnote{ 4$\times$3300 s on the first day and 4$\times$3000 s on the second day.}.

We extracted one-dimensional spectra using the M2FS IRAF package\footnote{https://ui.adsabs.harvard.edu/abs/ 2012SPIE.8446E..4YM/abstract}, which was described in  \cite{ref30}. In summary, we implemented CCD pre-processing, including overscan corrections, bias/dark subtraction, and cosmic ray removal. Then, four consecutive images of each night were combined. Next, multiple one-dimensional spectra were extracted using the IRAF package APALL \citep{ref31,ref32}. Finally, wavelength calibration was performed on the basis of the observed Ne--Hg--Ar--Xe arc lamp.

\section{ANALYSIS AND RESULTS} \label{sec:floats}
\subsection{Fundamental stellar parameters}

\begin{figure}
   \centering
   \includegraphics[width=\hsize]{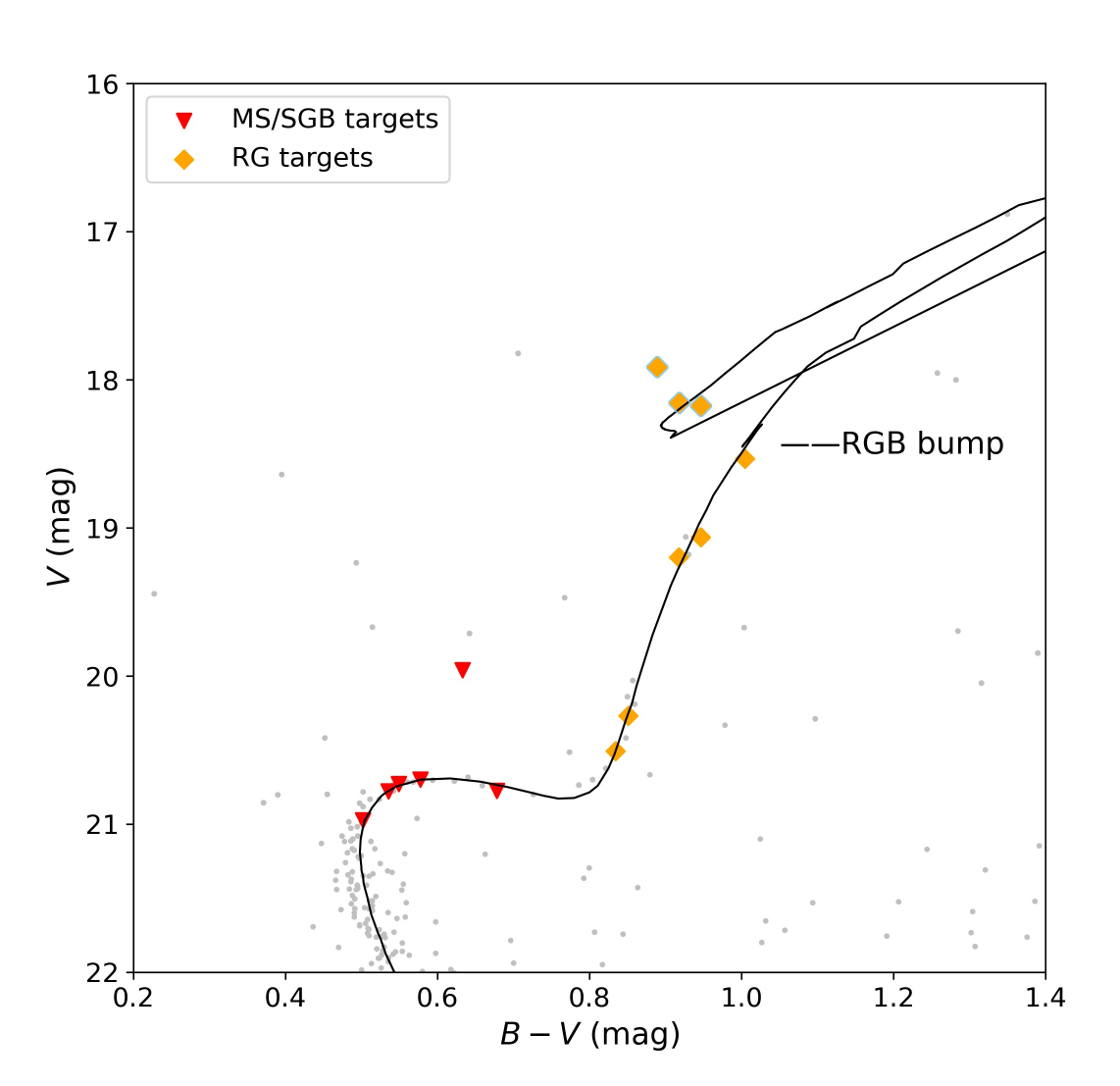}
      \caption{The color--magnitude diagram of Whiting 1. Red triangles and orange diamonds represent MS/SGB and RG stars in our high-confidence samples, respectively. Three possible red clump (RC) stars are marked with a blue edge. The gray dots are the stars within 1.1 $r_t$ of Whiting 1 in the Stetson photometry catalog. A PARSEC isochrone of age 5 Gyr and metallicity $-0.583$ is shown, and the position of the RGB bump is also indicated.}

         \label{isochrone}
   \end{figure}

To obtain high-confidence members of Whiting 1 after the arrival of Gaia DR3, we first performed a cross-match between the Whiting 1 photometry from the Stetson Photometric Catalog\footnote{https://www.canfar.net/storage/list/STETSON/Standards} and Gaia DR3 and used the following criteria to select members: (1) stars located within 1.1 $r_t$ (tidal radius), where $\rm r_t = 0.89 $ arcmin (\citealt{ref33}, 2010 edition); (2) proper motions (PMs) are consistent with the cluster PM $\rm (\mu_{\alpha cos\delta},\mu_\delta) = (-0.229,-2.047)mas\cdot yr^{-1}$\citep{ref34} within the uncertainties. Thus, we identified 44 member stars of Whiting 1 (shown as gray dots in Fig. \ref{isochrone}). We employ visual inspection to determine the best-fitting PARSEC isochrone for these stars. We varied the age from 5.0 to 6.5 Gyr in increments of 0.5 Gyr and Z from 0.003 to 0.005 in increments of 0.001. Our best estimated age is $\rm 5.0\pm0.5 Gyr$ with $\rm [Fe/H] = -0.58\pm0.05$. The distance modulus is $m-M = 17.84\pm 0.04$ ($\sim 31.6$ kpc), with a reddening of $\rm E(B-V) = 0.11\pm 0.01 $ mag. Among the selected stars, 14 were identified by our spectral observations: six subgiant branch (SGB) stars, five red giant branch (RGB) stars, and three possible RC stars (see Fig. \ref{isochrone}).

The effective temperature ($T_{eff}$) of each star in our high-confidence sample was calculated using the dereddened E(B-V) color, with the color--temperature conversion from  \cite{ref35}. For stars with temperatures higher than 5600 K, the CN and CH molecular lines are weaker and are considerably affected by Balmer lines. Therefore, in the subsequent analysis, we opted to exclude these hotter and fainter stars and focus solely on the eight brighter red giants, which form our final sample. Next, we calculated the surface gravity (log $g$) of each star using the relationship between $T_{eff}$ and log $g$ derived from the isochrone. The basic stellar parameters of the stars in our final sample are presented in Table 1. The errors in $T_{eff}$ and log $g$ were calculated by error propagation ($ \Delta T_{eff} \sim 120$ $ \rm K ,\Delta log$ $g \sim 0.1$).

To calibrate the flux of our final sample stars, we followed the work of \cite{ref17}, taking reference to their atmospheric model spectra. We used iSpec \citep{ref36} to generate a synthetic model spectrum of each star with the stellar parameters obtained previously. Here, we employed the line lists and radiative transfer code from SPECTRUM\footnote{http://www.appstate.edu/~grayro/spectrum/spectrum.html}\citep{ref37}. We applied the MARCS model atmospheres \citep{ref38} and solar abundances from  \cite{ref39}.
We then calibrated the observed spectrum with a synthetic spectrum by fitting a fourth-order polynomial to its continuum regions. Finally, the flux-calibrated spectra from the two nights were de-redshifted\footnote{using the cross-match correlation algorithm provided by iSpec.} and combined. The signal-to-noise ratio of our target stars is approximately 30.

\begin{table*}[t]
\footnotesize
\caption{Stellar parameters of the eight target stars}
\label{tab1}

\tabcolsep 6pt 
\begin{tabular*}{\textwidth}{ccccccccccccc}
\toprule
name &R.A.(deg) &Dec(deg) &B-V &V &$\mu_{\alpha\cos(\delta)}$ & $\mu_{\delta}$ &$T_{eff}$ (K) &log $g$&CN3839&CN3839e&CH4300&CH4300e     \\\hline
    B5-11 &30.734788 &--3.252114 &0.9464 &18.1748 &--0.450 &--2.092 &4970  &2.30 &0.143 & 0.017 & 0.342 & 0.009\\
    B5-12 &30.734082 &--3.248443 &0.8340 &20.5060 &0.492 &0.177 &5241  &3.39  &--0.014 & 0.066 & 0.281 & 0.025\\
    B5-13 &30.730941 &--3.244120 &0.9182 &18.1549 &--0.354 &--2.044 &5033  &2.37  & 0.130 & 0.023 & 0.332 & 0.006\\
    R3-07 &30.746751 &--3.242167 &0.9174 &19.1998 &0.148 &--2.151 &5035 &2.88  &0.102 & 0.033 & 0.313 &	0.011\\
    R5-03 &30.738546 &--3.264621 &0.8512 &20.2696 &--0.391 &--0.740 &5196  &3.28 & 0.004 & 0.047	& 0.280 & 0.011\\
    R5-05 &30.739998 &--3.254629 &0.9464 &19.0618 &--0.235 &--2.106 &4970  &2.73  &0.097 & 0.034 & 0.307 & 0.010\\
    R5-06 &30.739109 &--3.248986 &0.8892 &17.9129 &--0.354 &--1.997 &5101  &2.45  &0.154 & 0.018 & 0.297 & 0.004\\
    R5-12 &30.745510 &--3.261295 &1.0044 &18.5328 &--0.567 &--2.364 &4847   &2.49 &0.127 & 0.047 & 0.315 & 0.007\\
\bottomrule
\end{tabular*}
\end{table*}


\subsection{Spectral indices and abundances}

\begin{equation}
CN3839=-2.5log \frac{F_{3861-3884}}{F_{3894-3910}}
\label{eq:1}
\end{equation}

\begin{equation}
CH4300=-2.5log \frac{F_{4286-4315}}{0.5F_{4240-4280}+0.5F_{4390-4460}}
\label{eq:2}
\end{equation}

where $F_{X-Y}$ is the summed spectral flux from X to Y Å.

To analyze [C/Fe] and [N/Fe] specifically, we generated a set of model spectra covering [C/Fe] values from --0.5 to 0.3 ($\Delta$[C/Fe] = 0.02 dex) and [N/Fe] values from --0.5 to 1.0 ($\Delta$[N/Fe] = 0.1 dex) in iSpec for each star. Subsequently, we computed the model spectral indices of CN4142 and CH4300 for each of these spectra, constructing a grid of model spectral indices specific to each star. To find the best match to our observed spectral indices, we performed Newton’s first-order interpolation within the model grid. To estimate the measurement uncertainties in the C and N abundances, we ran Monte Carlo simulations with 5000 pairs of mock measurements for each star. The mock measurements were Gaussian distributions centered on the measured spectral indices with their errors as $\sigma$. Here, we considered different sources of errors, including flux calibration errors and errors propagated from stellar parameters ($T_{eff}$, log $g$, and [Fe/H]).
The resulting medians and standard deviations (errors) of [C/Fe] and [N/Fe] obtained by Monte Carlo simulations are presented in Table 2.

\begin{figure}
   \centering
   \includegraphics[width=\hsize]{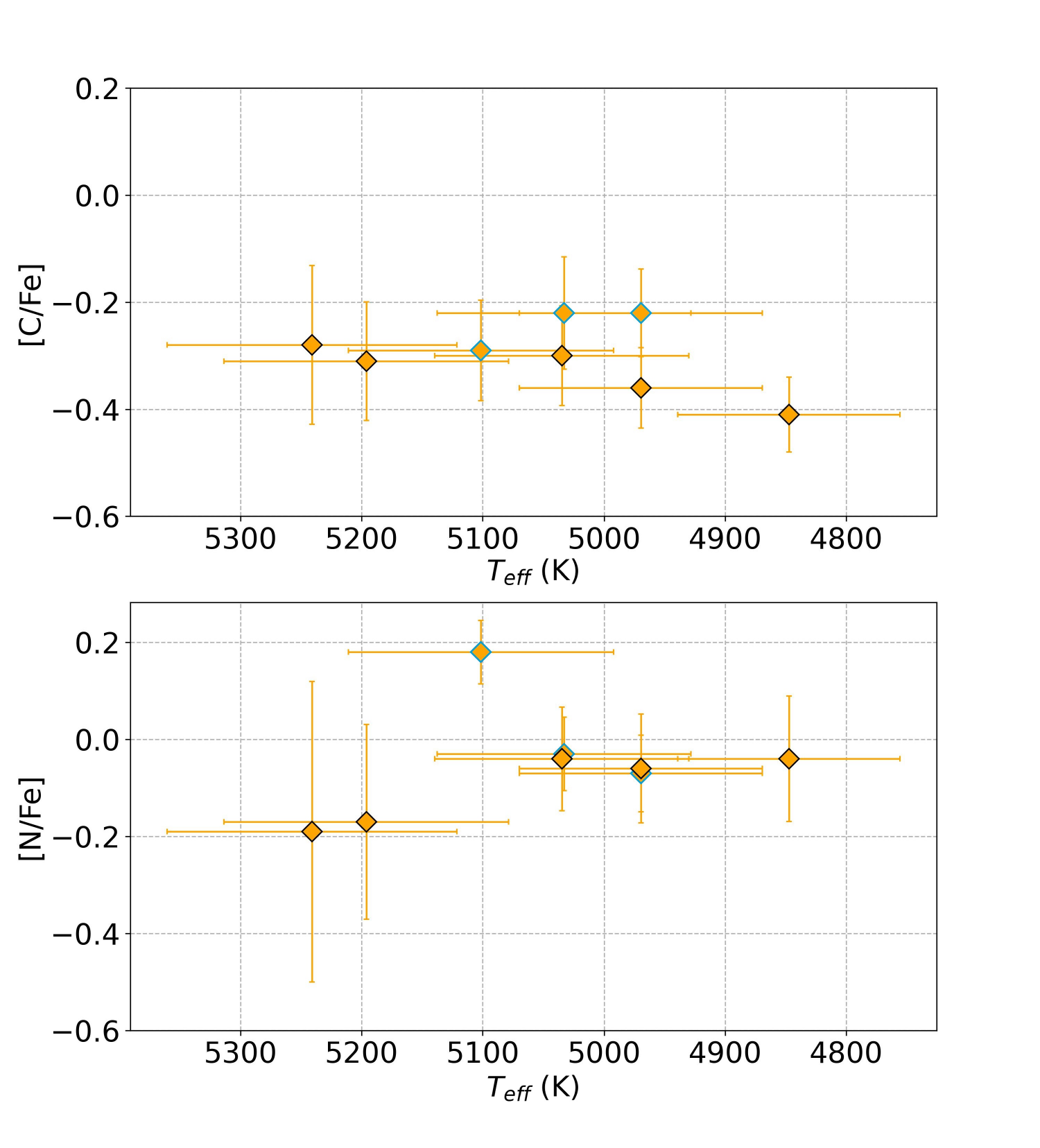}
      \caption{Distribution of the abundance ratios of [C/Fe] and [N/Fe]. Diamonds with blue edges represent possible RC stars.}
         \label{CNabundance}
   \end{figure}

An important indicator of the presence of MPs is that the scatter of C and N is larger than the observed uncertainties \citep{ref35,ref41,ref42}. Fig. \ref{CNabundance} shows the C and N abundances of our target stars as a function of $T_{eff}$. The standard deviation is approximately 0.06 dex for C and 0.10 dex for N. Considering that the mean errors of [C/Fe] and [N/Fe] are 0.09 dex and 0.13 dex, respectively, we do not detect abundance variation greater than the measurement errors for either element.
Our sample stars may have different levels of dredge-up and extra mixing during their RGB evolution, which is a concern. However, we do not observe an increase in [N/Fe] as a function of $T_{eff}$ in Figure \ref{CNabundance}. Because our RGB stars have all passed the first dredge-up and have not yet passed the RGB bump, they are probably affected approximately equally by the first dredge-up. In summary, the scatter and abundance errors of [C/Fe] and [N/Fe] are comparable. Therefore, we do not detect any MP signal in Whiting 1 based on our current observations.

Because we analyze only eight member stars in this study, we may miss any enriched population by chance. If we assume that the fraction of FG stars in Whiting 1 is 60\% \citep{ref43}, then the probability of seeing only one of the two populations based on a sample of eight stars is $P\sim{0.6}^8 = 0.0168 = 1.68\%$. This low probability strongly suggests that Whiting 1 is a genuine SSP.

\begin{table}[]
\centering
\caption{C and N Abundances of Observed Red Giants}\label{tab2}
\begin{tabular}{ccccc}
\toprule
 name &[C/Fe] &[C/Fe]err &[N/Fe] &[N/Fe]err     \\\hline
   B5-11&--0.22&0.08&--0.07&0.08     \\
   B5-12&--0.28&0.15&--0.19&0.31     \\
   B5-13&--0.22&0.10&--0.03&0.08     \\
   R3-07&--0.30&0.09&--0.04&0.11    \\
   R5-03&--0.31&0.11&--0.17&0.20     \\
   R5-05&--0.36&0.08&--0.06&0.11    \\
   R5-06&--0.29&0.09&0.18&0.07     \\
   R5-12&--0.41&0.07&--0.04&0.13     \\
\bottomrule
\end{tabular}

\end{table}

 \begin{figure*}
   \centering
   
   \includegraphics[width=0.9\textwidth]{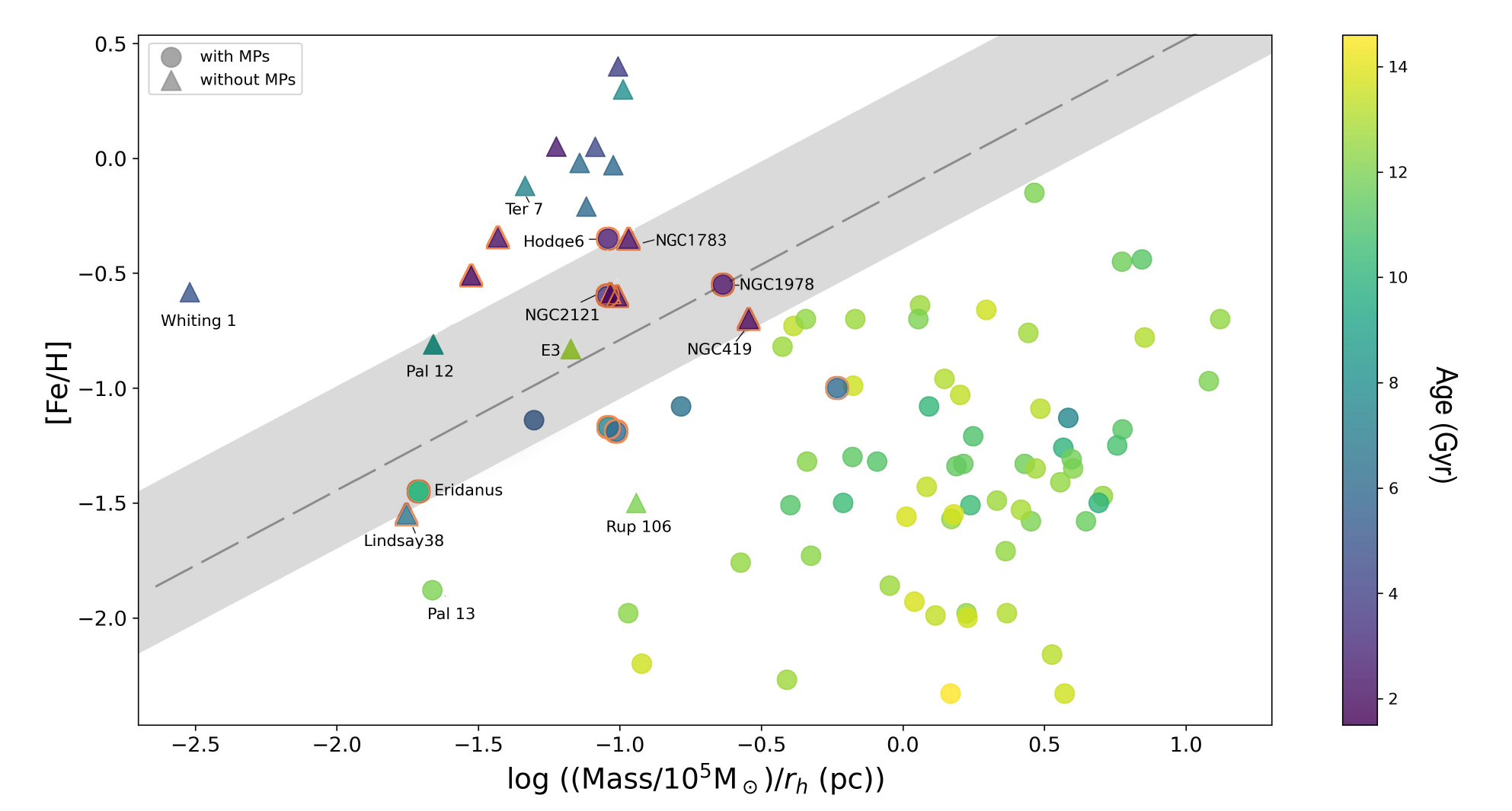}
      \caption{Logarithm of compactness versus metallicity diagram for clusters older than 1.5 Gyr. Cluster compactness is defined as the ratio of the initial stellar mass $\rm M_*$ to the half-mass radius $r_h$, scaled by $\rm 10^5 M_\odot$ and in parsecs (pc). Circles and triangles indicate clusters with and without MPs, respectively. Clusters with red edges are greater than 50 kpc from the Galactic center. Their ages are indicated by their colors. The dashed line represents the proposed limit between Galactic clusters with/without MPs (OCs/GCs). The shadow region represents the error after considering distant clusters.}
               \label{C5}
         
   \end{figure*}

\begin{table*}[t]
\footnotesize
\caption{Stellar parameters of Galactic star clusters}
\label{tab3}
\renewcommand\arraystretch{1}
\tabcolsep 12pt 
\begin{tabular*}{\textwidth}{cccccccccc}
\toprule

Object	  & Age(Gyr)  &	[Fe/H]    &	 MPs &	logC5  & Object	  & Age(Gyr)  &	[Fe/H]    &	 MPs &	logC5\\\hline
NGC104	  &  12.8	  &  --0.76	  &   Y	  &  0.44  &NGC6496	  &  12.4	  &--0.7	  &   Y	  &  --0.34\\
NGC288	  &  12.2	  &  --1.32	  &   Y	  &  --0.34&NGC6535	  &  10.5	  &--1.51	  &   Y	  &  0.24\\
NGC362	  &  10	      &  --1.26	  &   Y	  &  0.57  &NGC6553	  &  12	      &--0.15	  &   Y	  &  0.46\\
NGC1261	  &  10.24	  &  --1.08	  &   Y	  &  0.09  &NGC6541	  &  12.9	  &--1.53	  &   Y	  &  0.42\\
NGC1851	  &  7.64	  &  --1.13	  &   Y	  &  0.58  &NGC6584	  &  11.3	  &--1.3	  &   Y	  &  --0.18\\
NGC1904	  &  12	      &  --1.58	  &   Y	  &  0.45  &NGC6624	  &  12.5	  &--0.7	  &   Y	  &  1.12\\
NGC2298	  &  12.4	  &  --1.98	  &   Y	  &  0.22  &NGC6637	  &  13.1	  &--0.78	  &   Y	  &  0.85\\
NGC2808	  &  11.2	  &  --1.18	  &   Y	  &  0.78  &NGC6652	  &  12	      &--0.97	  &   Y	  &  1.08\\
NGC3201	  &  11.1	  &  --1.51	  &   Y	  &  --0.40&NGC6656	  &  12.7	  &--1.49	  &   Y	  &  0.33\\
NGC4590	  &  12.7	  &  --2.27	  &   Y	  &  --0.41&NGC6681	  &  12.8	  &--1.35	  &   Y	  &  0.47\\
NGC4833	  &  12.5	  &  --1.71	  &   Y	  &  0.36  &NGC6715	  &  10.8	  &--1.25	  &   Y	  &  0.76\\
NGC5024	  &  12.7	  &  --1.86	  &   Y	  &  --0.05&NGC6717	  &  13.2	  &--1.09	  &   Y	  &  0.49\\
NGC5053	  &  12.3	  &  --1.98	  &   Y	  &  --0.97&NGC6723	  &  13.1	  &--0.96	  &   Y	  &  0.15\\
NGC5272	  &  11.4	  &  --1.34	  &   Y	  &  0.19  &NGC6752	  &  13.8	  &--1.55	  &   Y	  &  0.18\\
NGC5286	  &  12.5	  &  --1.41	  &   Y	  &  0.56  &NGC6779	  &  13.7	  &--2	      &   Y	  &  0.23\\
NGC5466	  &  13.6	  &  --2.2	  &   Y   &	 --0.92&NGC6809	  &  13.8	  &--1.93	  &   Y	  &  0.04\\
NGC5897	  &  12.3	  &  --1.73	  &   Y	  &  --0.32&NGC6838	  &  12.7	  &--0.82	  &   Y	  &  --0.43\\
NGC5904	  &  11.5	  &  --1.33	  &   Y	  &  0.21  &NGC6934	  &  11.1	  &--1.32	  &   Y	  &  --0.09\\
NGC5927	  &  12.7	  &  --0.64	  &   Y	  &  0.06  &NGC6981	  &  10.9	  &--1.21	  &   Y	  &  0.25\\
NGC5986	  &  12.2	  &  --1.35	  &   Y	  &  0.60  &NGC7078	  &  13.6	  &--2.33	  &   Y	  &  0.57\\
NGC6093	  &  12.5	  &  --1.47	  &   Y	  &  0.71  &NGC7089	  &  11.8	  &--1.31	  &   Y	  &  0.60\\
NGC6101	  &  12.5	  &  --1.76	  &   Y	  &  --0.57&NGC7099   &  14.6	  &--2.33	  &   Y	  &  0.17\\
NGC6121	  &  13.1	  &  --1.98	  &   Y	  &  0.37  &Ton2	  &  12	      &--0.7	  &   Y	  &  0.05\\
NGC6139	  &  11.6	  &  --1.58	  &   Y	  &  0.65  &Rup106	  &  12	      &--1.5	  &   N	  &  --0.94\\
NGC6144	  &  13.8	  &  --1.56	  &   Y   &  0.01  &Ter7	  &  7.4	  &--0.12	  &   N	  &  --1.33\\
NGC6171	  &  13.4	  &  --1.03	  &   Y	  &  0.20  &Pal12	  &  8.6	  &--0.81	  &   N	  &  --1.66\\
NGC6205	  &  11.7	  &  --1.33	  &   Y	  &  0.43  &E3	      &  12.8	  &--0.83	  &   N	  &  --1.17\\
NGC6218	  &  13.4	  &  --1.43	  &   Y	  &  0.08  &ESO452SC11&  10	      &--1.5	  &   Y	  &  0.69\\
NGC6254	  &  12.4	  &  --1.57	  &   Y	  &  0.17  &Pal13	  &  12       &	--1.88    &   Y	  & --1.66\\
NGC6304	  &  13.6	  &  --0.66	  &   Y	  &  0.29  &Eridanus  &	10	      &--1.45	  &   Y   & --1.71\\
NGC6341	  &  13.2	  &  --2.16    &   Y  &  0.53  &Whiting 1 &	5         &--0.58     &	  N   &	--2.52\\
NGC6352	  &  12.7	  &  --0.7	  &   Y	  &  --0.17&Berkeley39&	6	      &--0.21	  &   N	  &--1.12\\
NGC6362	  &  13.6	  &  --0.99	  &   Y	  &  --0.18&Collinder261&6        &--0.03     &	  N	  &--1.02\\
NGC6366	  &  13.3	  &  --0.73	  &   Y	  &  --0.39&Messier67 & 4.5       &	0.05      &	  N   &--1.09\\
NGC6388	  &  11.7	  &  --0.45	  &   Y	  &  0.77  &NGC 188	  & 6.2	      &--0.02	  &   N	  &--1.14\\
NGC6397	  &  13.4	  &  --1.99	  &   Y	  &  0.12  &NGC 6253  &	4	      &0.4        &	  N	  &--1.01\\
NGC6441	  &  11.2	  &--0.44	  &   Y	  &  0.84  &NGC 6791  &	8	      &0.3        &	  N	  &--0.99\\
NGC 6819  &	2.5       &0.051	  &   N	  &  --1.22\\

\bottomrule
\end{tabular*}
\end{table*}

\section{Discussion}\label{sec:4}
\subsection{Possible parameters distinguishing between star clusters with and without MPs}
Whiting 1 is a star cluster whose mass falls within the overlap region between massive OCs and low-mass GCs. Its initial mass is approximately $\rm 5\times 10^3 M_\odot$ \citep{ref44}. Its low mass could explain why Whiting 1 does not have MPs. The initial mass threshold for star clusters (older than 2 Gyr) to host MPs is between $\rm{log}M_{initial}\sim$ 4.98 and 5.26, according to \cite{ref45}. In this sense, the initial mass of Whiting 1 lies safely below the proposed mass threshold and is thus expected to agree with our findings.

However, the limit between star clusters with MPs and without MPs is not clear-cut in the initial mass versus age parameter space (e.g., Figure 8 of \citealt{ref45}). After exploring several sets of parameter space, we find that the aforementioned separation is most robust in the cluster compactness vs. metallicity space (Fig. \ref{C5}). We define cluster compactness as the initial mass divided by half the mass radius (scaled by $\rm 10^5 M_\odot$ and in units of parsecs), which is denoted as $C_5$ in \cite{ref46}. The cluster present-day masses and half-mass radii were determined by fitting N-body models to archival Hubble Space Telescope photometry \citep{ref47}. The initial masses of these clusters are calculated using the present-day masses and applying the models of  \cite{ref48}. The metallicities and ages of these GCs were taken from \cite{ref49} and \cite{ref50}, respectively.

If only Galactic star clusters with Galactic centric distances less than 50 kpc are considered (symbols not shown in red), we see a clear separation between clusters with and without MPs (GCs/OCs). Therefore, we draw an empirical line (depicted as a dashed line in Fig. \ref{C5}) to differentiate Galactic clusters with and without MPs. However, when considering clusters located at distances greater than 50 kpc, including clusters in the Large Magellanic Cloud (LMC) and Small Magellanic Cloud (SMC), their metallicities and compactness are less well-constrained, and whether they have MPs is still under debate.
The ages of the SMC/LMC clusters were obtained from \cite{ref51}. Metallicity measurements may suffer from large uncertainties because of their remoteness. Therefore, the metallicity of each LMC/SMC cluster is averaged using data from several literature sources (refer to Table 4 for details).
The large uncertainties cause the mixing of two types of star clusters near the limit, and we include a shadowed stripe to account for the potential observation errors in these cases. The most anomalous cluster in our classification scheme is Rup 106, which, although relatively low in mass and compactness, is quite metal-poor and clearly lies in the regime otherwise occupied by clusters with MPs but surprisingly does not possess MPs\citep{ref52}.

What is the physics between this empirical line/region?  \cite{ref46} proposed a convincing model:
Less compact star clusters could not resist the gas expulsion of the primordial population, leading to insufficient enriched gas retained for any subsequent star formation, thus suppressing the MP phenomena. In contrast, the enriched gas is retained by clusters with higher densities and allowed to accumulate within these clusters. The higher the metallicity, the stronger the stellar winds and the less likely MPs will be produced. Combining these two factors, clusters with low density and high metallicity are prone to strong gas expulsion, resulting in the inability to accumulate gas within the cluster and subsequently failing to form MPs.

In the top left of Fig. \ref{C5}, we focus on four star clusters without MPs: Whiting 1, Terzan 7 \citep{ref53}, Palomar 12 \citep{ref54}, and E3 \citep{ref55}, which are traditionally classified as GCs based on their older ages and high Galactic latitudes. However, they lack the characteristic features of GCs: MPs. 
Interestingly, all these star clusters have ex-situ origins. Whiting 1, Terzan 7, and Palomar 12 are connected to the Sgr tidal streams. These clusters share the same age--metallicity relationship as stars in Sgr dSph. Additionally, E3 is another sparse star cluster with a possible extragalactic origin \citep{ref56}.

This observation may be related to the galactic environment in which they were formed: Each galaxy has its own age--metallicity relation for star clusters \citep{ref56}. GCs born in situ are mostly old (age $ > $ 10 Gyr) and compact because of the dense, metal-poor galactic environment during their formation. After the proto-galaxy phase, the MW Galactic environment is no longer suitable for GC formation. More metal-rich OCs were formed after the formation of a stellar disk. In contrast, dwarf galaxies are less compact because of their low gravitational potential, and their star cluster formation could be vastly different from that of the MW. Star clusters formed during later starbursts in dwarf galaxies may exhibit distinct compactness and metallicities. Meanwhile, the formation of their internal stellar populations would be influenced by the tidal field of their host galaxies \citep{ref57}. The absence of MPs in these four ex situ star clusters may not be a coincidence, as metal-rich, diffuse, old ($ > $ 2 Gyr) star clusters have a higher survival rate if they come from satellite dwarf galaxies.

\subsection{Sgr dwarf galaxy and its associated star clusters}

The stars in galaxies are suggested to be born in star clusters; therefore, star cluster properties would substantially affect galaxy formation. Our recently proposed project ``Scrutinizing {\bf GA}laxy-{\bf ST}a{\bf R} cluster coevoluti{\bf ON} with cheom{\bf O}dyna{\bf MI}cs ({\bf GASTRONOMI})'' aims to investigate the interplay between MW, nearby satellite dwarf galaxies, and star clusters using photometric and spectroscopic data. This project bridges star evolution and galaxy evolution with solid foundations, which may also help us understand exotic objects in the early universe.

 Sgr has undergone multiple star formations, and the members follow a unique age and metallicity relation \citep{ref58,ref59,ref60,ref61}.
The oldest star formation episode dates back to at least 10 Gyr ago ($\rm [Fe/H] < -1.3$), while multiple more recent star formation events generated intermediate ($\rm 4-6 \, Gyr, [Fe/H]\approx -0.9\sim -0.5$) and young populations ($\rm 0.5\sim 2 \, Gyr,[Fe/H] > -0.4$). The ages of the intermediate and young populations nicely coincide with the recent star formation of the MW disk \citep{ref62}, suggesting that these epochs of star formation were stimulated by multiple passages of the Sgr dwarf galaxy. Along with the infall of Sgr, several star clusters were born, including Terzan 7, Palomar 12, Whiting 1, and Berkeley 29. Thus, they considerably differ from star clusters that were formed earlier ($ > 10$ Gyr): They are more metal-rich, less compact, and have distinctive horizontal branch morphology \citep{ref25}. 
In addition, they do not have the characteristic features of GCs: MPs. 
We depict the interplay between MW, Sgr, and their associated star clusters as follows: (1) Metal-poor, old GCs (with MPs) were formed separately in MW and Sgr during the early phase ($ > 8-10$ Gyr ago); (2) multiple passages of the Sgr dwarf galaxy stimulated epochs of star formation in MW and Sgr. As the enriched galactic environment is no longer suitable for MP formation, more metal-rich, less compact star clusters without MPs were formed during this phase. The later-formed star clusters in Sgr may survive longer than the MW OCs in the absence of a galactic disk; and (3) star clusters in MW and Sgr were mixed spatially. The newly discovered star clusters associated with Sgr \citep{ref63, ref64,ref65,ref66} may verify this scheme after a more detailed investigation is available.

If we follow the definition of  \cite{ref13}, then $bona fide$ GCs are ``stellar aggregates showing the Na-O anticorrelation'' or, in other words, MPs, and we may reach another interesting conclusion. The diffuse star clusters without MPs, i.e., Whiting 1, Terzan 7, Palomar 12, and E3, are not GCs but OCs. Furthermore, the later-formed star clusters in Sgr, including Whiting 1, Terzan 7, Palomar 12, and Berkeley 29, are OCs of Sgr. Interestingly, these four star clusters show typical horizontal branch morphology similar to other OCs; only the RC is visible. This conclusion could considerably impact the galactic evolution studies of the Sgr dwarf galaxy.

\section{Conclusions}\label{sec:5}
The search for physical parameters that separate star clusters with and without MPs has been the ``holy grail'' for the study of star cluster formation. We are currently undertaking an observational program to search for MP signals in old star clusters that linger between GCs and OCs. In this study, we analyzed the low-resolution spectra of eight stars in the cluster Whiting~1, which has a current mass of only $\rm 2\times 10^3 M_{\odot}$. We measured the CN3839 and CH4300 spectral indices and evaluated the corresponding C and N abundances by comparing the stellar indices. The C and N abundances do not show substantial internal variations compared with measurement errors, indicating an absence of MPs in Whiting 1.

 Combining our results with literature studies of MPs, we found a possible limit between clusters with/without MPs in the metallicity vs. compactness index parameter space. This result qualitatively agrees with the gas expulsion model described by  \cite{ref46}: Diffuse clusters with high metallicity are less likely to maintain the enriched gas and thus fail to generate the SG required to display MPs.

The galactic environment has an important influence on the nature of star clusters formed within a galaxy. Metal-rich, diffuse, old star clusters have a higher survival rate if they were formed in dwarf galaxies because their passages through dense galactic components, e.g., a galactic disk, are minimized. Our proposed limit also considerably impacts the Sgr star cluster formation picture: Younger star clusters formed after the first starburst epoch ($ > 8-10$ Gyr), along with the infall of Sgr, do not have the characteristic features of GCs and MPs, including at least Whiting 1, Terzan 7, Palomar 12, and Berkeley 29.

\begin{table}
\centering
\caption{Stellar parameters of LMC/SMC star clusters}\label{tab4}

\begin{tabular}{ccccc}
\toprule
Object	                       & Age(Gyr)&[Fe/H]    &MPs &logC5  \\\hline
NGC1978\citeyearpar{ref67}	           & 2	     & --0.55	& Y	 & --0.64\\
NGC121\citeyearpar{ref68}	           & 10.5	 & --1.5	& Y	 & --0.21\\
Lindsay1\citeyearpar{ref69,ref70}     & 7.5	 & --1.17	& Y	 & --1.04\\
NGC339\citeyearpar{ref69}	           & 6	     & --1.19	& Y	 & --1.01\\
NGC416\citeyearpar{ref71}	           & 6	     & --1	    & Y	 & --0.23\\
Kron3\citeyearpar{ref72}	           & 6.5	 & --1.08	& Y	 & --0.78\\
NGC2121\citeyearpar{ref73}	           & 3.2	 & --0.6	& Y	 & --1.04\\
Hodge6\citeyearpar{ref70,ref74,ref75} & 2.5	 & --0.35	& Y	 & --1.04\\
NGC1806\citeyearpar{ref76}	           & 1.5	 & --0.6	& N	 & --1.01\\
NGC1846\citeyearpar{ref77}	           & 1.5	 & --0.59	& N	 & --1.03\\
Lindsay38\citeyearpar{ref78,ref79}	   & 6.5	 & --1.55	& N	 & --1.75\\
Lindsay113\citeyearpar{ref69,ref80}   & 5.3	 & --1.14	& N	 & --1.30\\
NGC1651	\citeyearpar{ref81}           & 2	     & --0.3	& N	 & --1.43\\
NGC1783	\citeyearpar{ref81}           & 1.8	 & --0.35	& N	 & --0.97\\
NGC2173\citeyearpar{ref81}	           & 1.6	 & --0.51	& N	 & --1.52\\
NGC419\citeyearpar{ref82}	           & 1.5	 & --0.7	& N	 & --0.54\\
\bottomrule
\end{tabular}

\end{table}

\begin{acknowledgments}
The authors would like to express gratitude to Jing Zhong and Li Chen for their constructive comments. Additionally, two anonymous reviewers and the editor provided valuable feedback, which significantly contributed to the improvement of the manuscript. Ruoyun Huang, Baitian Tang, and Chengyuan Li gratefully acknowledge support from the Natural Science Foundation of Guangdong Province under grant No. 2022A1515010732, the National Natural Science Foundation of China through grants No. 12233013, and the China Manned Space Project Nos. CMS-CSST-2021-B03, CMS-CSST-2021-A08, and etc. Chengyuan Li also acknowledges financial support from National Natural Science Foundation of China through grants No. 12073090. Doug Geisler gratefully acknowledges the support provided by Fondecyt regular n. 1220264. Doug Geisler also acknowledges financial support from the Direcci\'on de Investigaci\'on y Desarrollo de la Universidad de La Serena through the Programa de Incentivo a la Investigaci\'on de Acad\'emicos (PIA-DIDULS). JAC-B acknowledges support from FONDECYT Regular N 1220083. Bruno Dias acknowledges support by ANID-FONDECYT iniciación grant No. 11221366. Jos\'e G. Fern\'andez-Trincado F-T gratefully acknowledges the grant support provided by Proyecto Fondecyt Iniciaci\'on No. 11220340, and also from the Joint Committee ESO-Government of Chile 2021 (ORP 023/2021).
\end{acknowledgments}

%

\vspace{5mm}



\bibliography{main}{}
\bibliographystyle{aasjournal}

\end{document}